# Full Spectrum Diffused and Beamed Solar Energy Application Using Optical Fibre

M. R. Dutta Majumdar and Debasish Das
Variable Energy Cyclotron Centre,1/AF, Bidhan Nagar, Kolkata – 700 064, INDIA
email: mrdm@veccal.ernet.in

*ABSTRACT:* Existing solar energy application systems use small fraction of full spectrum of solar energy. So attempts are made to show how full spectrum solar energy can be used for diffused and beamed form of incident solar energy. Luminescent Solar Concentrator (LSC) principle with optical fibre in diffused sun light and dielectric mirror separation technique with optical fibre in beamed form are discussed. Comparison of both the cases are done.

## I. INTRODUCTION

Solar photonics, which is an emerging area, has culminated the use of photonics, optoelectronic components and light-wave propagation media towards the successful use of solar energy.

Solar energy incident on earth surface has wide spectrum ranging from 200 nm to 2500 nm. The entire spectrum is divided into few wave-length regions. The major application areas of solar energy are photosynthesis, illumination, heating and photovoltaic process.

Solar radiation reaching earth surface may be in diffused or directional beam form. So concentrated collection of solar radiation needs lens-mirror combination for beamed radiation and Luminescent Solar Concentrator (LSC) principle [1,2] for diffused radiation.

Some attempts are done to propose schemes for better and efficient way of utilization of solar energy in daily life.

## II. METHODOLOGY

### A. Solar energy application limitation

Solar energy available in various regions is typically 8.3% ultra-violet (200 nm –400 nm), 38.2% visible (400 nm – 700 nm), 28.1% near infra-red (700 nm – 1100 nm) and 25.4% infrared/far-infrared portion.

All other processes excluding heating require a small portion of solar spectrum. In case of heating near infrared and infrared portions are more useful and efficient. Photo-synthesis requires very selective bands of radiation in visible and near infrared region.

Illumination requires radiation in the range of 400 nm to 700 nm. Silicon Photo Voltaic (PV) cells efficiency is nearly 10% and are not sensitive above 1100 nm. Over all efficiency is reasonable within the region of 700 nm to 1100 nm.

One can see that for a solar irradiated area, only a small amount of incident radiation is used out of full spectrum solar energy. But by the proposed full spectrum application schemes one can use simultaneously almost all radiation for different ways of application.

### B. Collection of diffused solar energy

Any imaging optical device may not possibly concentrate diffused solar radiation coming to earth's surface due to clouds and dust. The alternative method of light collection under such conditions is the use of luminescent (fluorescent) or Wave Length Shifting (WLS) radiation converters.

A luminescent solar concentrator is essentially a large transparent area of high refractive index material doped with suitable inorganic or organic dopant material and surrounded by low refractive index material.

A portion of the solar spectrum is absorbed in the dopant and then re-emitted isotropically at longer wave-length where it is weakly re-absorbed. Small fraction of this light is trapped inside the transparent medium due to total internal reflection occurring at the boundary of denser and lighter media. Light trapped inside sheet material comes out at the edges.

Production of white light by mixing outputs of LSC plastic sheets [3,4] of three fundamental color like Red, Blue and Green were reported earlier. We also present how LSC technique may be useful using luminescent optical fibre for light trapping [5] and further studies done with purely diffused light for multiple applications.

## C. Normal - Luminescent optical fibre:

Normal optical fibre: In normal optical fibre light launched along the axis of optical fibre to core (high refractive index) with suitable angle experience internal reflection at the core, clad (low refractive index) interface. This way light propagates to distant place as seen in communication optical fibres.

But for solar energy light transmission optical fibre should be of low loss, high numerical aperture and large cross-section . It should transmit wide spectrum of solar radiation at high intensity without degradation of optical property of optical fibre material. Some applications of solar energy with optical fibre are:
- Optical fibre for solar energy transport [6]
- Plane of mini-dish collection system [7]
- Intense solar thermal energy [8] for turbine
- Hybrid day lighting [9]

Materials for normal optical fibre for solar energy work are:
- Pure quartz glass for ultra violet region
- Plastic optical fibre for visible region
- Silica glass for near IR or IR region
- Halide glass or Hollow Wave Guide(HWG) for IR and far IR. region

Luminescent optical fibre: The core material is mixed with fluorescent/luminescent material. A small fraction of light trapped inside core due to total internal reflection and transmitted to both ends of the fibre. Diffused light can be collected and the magnitude of light collection depends on trapping efficiency, absorption emission spectra of luminescent dye and exposer length as explained in our earlier work [5]. We have used luminescent fibre for High Energy Physics experiments [10] and developed splicing jigs for clear and luminescent fibres. We have used three color luminescent fibre of thin cross-section typically about .25 mm and spliced with clear (undoped) 1 mm plastic fibre. This way of splicing makes exit angle of light smaller into clear fibre region. Light trapped inside luminescent fibre experience high trapping efficiency due to high numerical aperture as light is trapped in core to air interface.

## D. Diffused Light Collection schemes

This scheme involves following steps:

a) A setup shown is in Fig (1) to measure light output of three colors (RBG) luminescent fibre with change of exposer length. Measurements of light outputs were done with a set of calibrated photodiodes by measuring photo current. Other ends of three fibres are connected for light mixing and white light is produced. Mixed light can be compared with standard light source (white LED) or with three colors (RBG) LED light source. Other than comparison of lights in setup there is provision to measure light illumination level by setting photometric photo-diodes and measuring photo-current. This photometric photo-diodes response is similar to day light adapted [3] human eye. The whiteness of light depends on chromatic diagram as discussed in ref. [3].

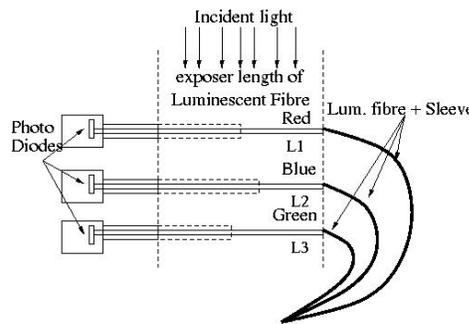

Fig 1(a)

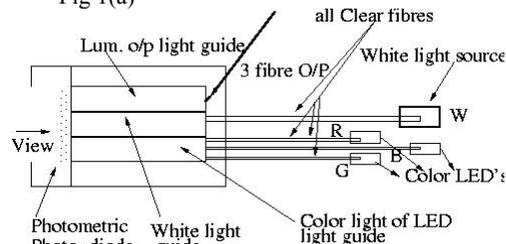

Fig 1(b)

b) Multiple application setup:

A setup shown in Fig (2) presents the multi-functional application of diffused light in three different ways like (i) photo-voltaic power generation, (ii) white light production and (iii) heat collection.

Schematic figure shows:
- Thin luminescent fibres are thermally spliced with thick clear plastic fibres in divergent

tapering way to reduce exit end numerical aperture as shown in Fig 2(a).

- Formation of three color (RBG) luminescent fibre layer separated by air and enclosed in transparent box as in Fig 2(b).
- One end of three optical fibre bundles joined with light guide for production of white light as seen in left side of Fig 2(c).
- Another end of three optical fibre bundles joined with set of PV cells for electric power generation as seen in right side of Fig 2(c).
- Remaining unabsorbed part mainly as infrared goes for thermal heating for heating of fluid like water as seen in bottom part.

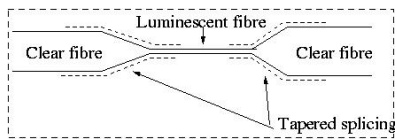
Fig 2(a)

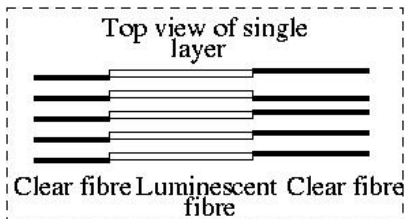
Fig 2(b)

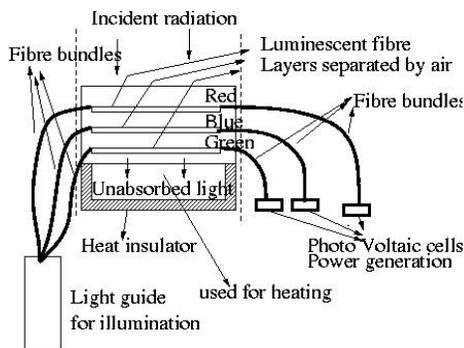
Fig 2(c)

Estimates are shown in result

### E. Collection of Beamed solar energy

In case of beamed solar energy maximum realizable flux concentration in air is $C_{max}$, often referred to as the thermodynamic limit as it is derived [11] from the second law of thermodynamics:

$C_{max} = 1/\sin^2(\theta_s) = 45{,}000$ sun intensity.

where Solar disk subtending a half-angle $\theta_s$ of $0.27^0$ at the earth. Flux density (one sun) of only 137 mw/sq.cm extra-terrestrially is available. The clear-sky direct beam radiation typical average energy is 110 mw/sq.cm at earth's surface. Solar energy illumination efficacy is 110 lumens/watt. Collection of that radiation is possible with imaging optical device like lens and mirror combination for concentrating to a high intensity.

Now with above level of concentrated solar energy this may be useful for application using suitable dielectric material devices. Separation of wave-lengths like visible region, near infra-red and far infrared is possible using wave-length selective dielectric mirrors.

### F. Beamed Radiation scheme

Beamed solar radiation collection scheme of full spectrum solar energy application shown in fig 3 involves:

- Collection of solar radiation using parabolic mirror.
- Separating visible, near infrared and infrared by wavelength selective dielectric mirrors.
- Separated radiation components may be focused to three different sets of optical fibres where material choice is discussed earlier.
- Visible region is useful to remote daylight illumination.
- Near infrared region could be utilized with conventional silicon solar cells at medium concentration with higher efficiency with low heat loss.
- Infrared part may be transported for remote heating of a system or Thermo Photo Voltaic (TPV) [9] power generation using low band gap material.
- Estimates are shown in result.

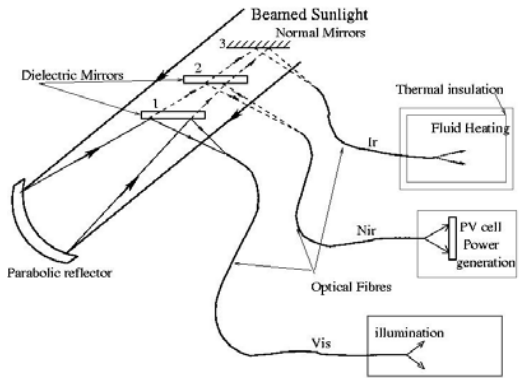

Fig 3

## III. DATA AND RESULTS
### A. Luminescent fibre measurement

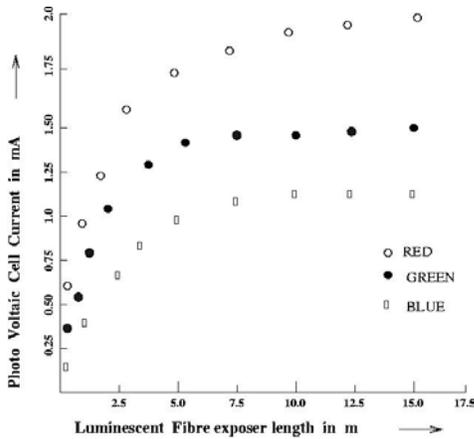

Fig 4 Length vs Output: output of exposed luminescent fibre measured by photo diodes for each luminescent fibre with typical open sky sunlight exposer.

### B. Estimates on Diffused light:
Table 1

| Application | Power output | % efficiency |
|---|---|---|
| PV power | 15 mw | 1.50 |
| illumination | 20 lumen/watt | 18.2 |
| Heating | 300 mw | 30.0 |
| Total | 497 | 49.7 |

### C. Estimates on Beamed light:
Table 2

| Application | Power output | % efficiency |
|---|---|---|
| PV power | 63 mw | 6.3 |
| illumination | 55 lumens/watt | 50.0 |
| Heating | 190 mw | 19.0 |
| Total | 753 mw | 75.3 |

## Conclusions

- Our observation with both the schemes: Diffused way of collection and application is always less efficient than beamed light collection. But some places diffused way collection may be important.
- Beamed schemed is based on our conservative estimates only. But after some real experiment we feel this scheme have huge potential for large amount of solar energy applications.

## ACKNOWLEDGEMENTS

We gratefully acknowledge Dr Bikash Sinha Director, VECC for his encouragement and support to this work. We also acknowledge Dr Y. P. Viyogi, now Director of Institute of Physics, Bhubaneswar and Dr T. K. Nayak, VECC for their encouragement and valuable discussions. We further acknowledge our high energy physics colleagues for making valuable suggestions time to time.